\documentstyle[graphicx,12pt]{article}
\begin{document}

\small
\hoffset=-1truecm
\voffset=-2truecm
\title{\bf The asymptotic behavior of Casimir force in the presence of compactified universal extra dimensions}
\author{Hongbo Cheng\footnote {E-mail address:
hbcheng@public4.sta.net.cn}\\
Department of Physics, East China University of Science and
Technology,\\ Shanghai 200237, China}

\date{}
\maketitle

\begin{abstract}
The Casimir effect for parallel plates in the presence of
compactified universal extra dimensions within the frame of
Kaluza-Klein theory is analyzed. Having regularized and discussed
the expressions of Casimir force in the limit, we show that the
nature of Casimir force is repulsive if the distance between the
plates is large enough and the higher-dimensional spacetime is,
the greater the value of repulsive Casimir force between plates
is. The repulsive nature of the force is not consistent with the
experimental phenomena.
\end{abstract}
\vspace{8cm} \hspace{1cm}
PACS number(s): 11.10.Kk, 04.62.+v

\newpage

Unifying the interactions in nature needs a powerful ingredient
like the model of higher dimensional spacetime. Nearly 80 years
ago the idea that our universe has more than four dimensions was
put forward by Kaluza and Klein [1, 2]. In this theory named
Kaluza-Klein theory, one extra dimension in our Universe was
introduced to be compactified in order to unify gravity and
classical electrodynamics. Recently the quantum gravity such as
string theories or brane-world scenario is developed to reconcile
the quantum mechanics and gravity with the help of introducing
seven extra spatial dimensions. In Randall-Sundrum model the
matter fields may be localized on a four-dimensional brane
considered as our real universe, and only gravitons can propagate
in the extra space transverse to the brane [3, 4]. In some
approaches larger extra dimensions were also invoked for providing
a breakthrough of hierarchy problem [5-7]. The order of the
compactification scale of the extra dimensions has not been
confirmed and are also of considerable interest recently. In a
word, studies of higher-dimensional spacetime have therefore been
pursed vigorously and extensively and more achievements have been
made.

The Casimir effect as a fundamental aspect of quantum field theory
in confined geometries and the physical manifestation of
zero-point energy has received great attention and has been
extensively studied in a wide variety of topics [8-18]. The topics
include the influence from the effect on the stability of radion
in the Randall-Sundrum model, the cosmological aspects like the
cosmological constant and the primordial cosmic inflation [19,
20]. The effect was also explored in the context of string theory
[21-24]. The precision of the measurement has been greatly
improved practically [25-28], leading the Casimir effect to be a
remarkable observable and trustworthy consequence of the existence
of quantum fluctuations. The experimental results clearly show
that the attractive Casimir force between the parallel plates
vanishes when the plates move apart from each other to the very
distant place. In particular it must be pointed out that no
repulsive force appears. Therefore the Casimir effect can become a
powerful tool for the study and development of a large class of
topics on the model of Universe with more than four dimensions.

Exploring the possible existence or size of extra dimensions by
means of Casimir effect attracts more attentions of the physical
community. Research on the Casimir effect in five-dimensional
spacetimes is just the first step of generalization to investigate
the higher dimensional spacetimes. Having examined the Casimir
effect for the rectangular cavity in the presence of a
compactified universal extra dimension, we show analytically that
the extra-dimension corrections to the standard Casimir effect are
very manifest [29]. The Casimir effect for parallel plates in the
spacetime with one extra compactified dimension was discussed.
Only when the plates gap is very small, the size of the additional
dimension satisfying $L\leq 10nm$ was obtained by comparison to
experimental data [30]. We also scrutinized the same problem and
show rigorously that there must appear repulsive Casimir force
between the parallel plates within the experimental reach when the
plates distance is large enough in the spacetime with one
compactified additional dimension [31]. Therefore the results
obtained from the Kaluza-Klein theory including only one
compactified spatial extra dimension are not consistent with the
experimental results mentioned above, which means that the model
that the spacetime with only one extra dimension can not be
realistic.

As mentioned above a lot of models such as the string theories
motivate the models with more than five dimensions, suggesting it
is necessary to continue exploring the Casimir effect in the
presence of more compactified universal extra dimensions in detail
in order to know whether the models of spacetime with more than
one additional dimensions are realistic. This problem, to our
knowledge, has not been discussed. For simplicity and comparison
to the measurement the system consisting of two parallel plates is
always chosen. The purpose of this paper is to reexamine the
Casimir effect for parallel plates in the universe with $d$
compactified spatial dimensions carefully. We regularize the total
energy to obtain the Casimir energy, and then Casimir force. In
particular we focus on the asymptotic behaviour of the Casimir
force between plates for their large enough gap and the dependence
of dimensionality of the spacetime in order to compare our results
with the measuring evidence listed above directly. Finally the
conclusions are emphasized.

In the Kaluza-Klein theory we start to consider the scalar field
in the system consisting of two parallel plates in the spacetime
with $d$ extra compactified dimensions. Along the extra dimensions
the wave vectors of the field have the form
$k_{i}=\frac{n_{i}}{L}$, $i=1, 2, \cdot\cdot\cdot, d$
respectively, $n_{i}$ an integer. Here we choose that the extra
dimensions possess the same radius as $L$. At the plates the
fields satisfy the Dirichlet condition, leading the wave vector in
the directions restricted by the plates to be
$k_{n}=\frac{n\pi}{R}$, $n$ a positive integer and $R$ the
separation of the plates. Under these conditions, the zero-point
fluctuations of the fields can give rise to observable Casimir
forces.

In the case of $d$ additional compactified dimensions we find the
frequency of the vacuum fluctuations to be,

\begin{equation}
\omega_{\{n_{i}\}n}=\sqrt{k^{2}+\frac{n^{2}\pi^{2}}{R^{2}}+\sum_{i=1}^{d}\frac{n_{i}^{2}}{L^{2}}}
\end{equation}

\noindent where

\begin{equation}
k^{2}=k_{1}^{2}+k_{2}^{2}
\end{equation}

\noindent $k_{1}$ and $k_{2}$ are the wave vectors in directions
of the unbound space coordinates parallel to the plates surface.
Here $\{n_{i}\}$ represents a short notation of $n_{1}, n_{2},
\cdot\cdot\cdot, n_{d}$, $n_{i}$ a nonnegative integer. Following
Refs. [9-16], therefore the total energy density of the fields in
the interior of system reads,

\begin{eqnarray}
\varepsilon=\int\frac{d^{2}k}{(2\pi)^{2}}\sum_{n=1}^{\infty}\sum_{\{n_{i}\}=0}^{\infty}\frac{1}{2}\omega_{\{n_{i}\}n}
\hspace{7.8cm}\nonumber\\
=\frac{\pi}{2}\frac{\Gamma(-\frac{3}{2})}{\Gamma(-\frac{1}{2})}\sum_{l=0}^{d-1}\left(%
\begin{array}{c}
  d \\
  l \\
\end{array}%
\right)E_{d-l+1}(\frac{\pi^{2}}{R^{2}}, \frac{1}{L^{2}},
\frac{1}{L^{2}},\cdot\cdot\cdot, \frac{1}{L^{2}};
-\frac{3}{2})+\frac{\pi^{3}}{2R^{3}}\frac{\Gamma(-\frac{3}{2})\zeta(-3)}{\Gamma(-\frac{1}{2})}
\end{eqnarray}

\noindent in terms of the Epstein zeta function $E_{p}(a_{1},
a_{2}, \cdot\cdot\cdot, a_{p}; s)$ defined as,

\begin{equation}
E_{p}(a_{1},a_{2},\cdot\cdot\cdot,a_{p};s)=\sum_{\{n_{j}\}=1}^{\infty}(\sum_{j=1}^{p}a_{j}n_{j}^{2})^{-s}
\end{equation}

\noindent here $\{n_{j}\}$ stands for a short notation of
$n_{1},n_{2},\cdot\cdot\cdot,n_{p}$, $n_{j}$ a positive integer.
By regularizing Eq.(2) and doing burden calculation, we can obtain
the Casimir energy density of parallel plates in the spacetime
with $d$ extra compactified spatial dimensions. It is certainly
fundamental to investigate the Casimir force for the same system
in the same background in order to compare our results with the
experimental phenomenon. The Casimir force is given by the
derivative of the Casimir energy with respect to the plate
distance. Here we focus on the property as the plate distance $R$
approaches to the infinity, then the expression for the Casimir
force in the limiting case is defined as,

\begin{equation}
f=-\lim_{\mu\longrightarrow\infty}\frac{\partial\varepsilon}{\partial
R}
\end{equation}

\noindent where

\begin{equation}
\mu=\frac{R}{L}
\end{equation}

\noindent Substituting Eq.(2) into (4) and proceeding the burden
derivation, we obtain the expression of the asymptotic behavior of
the Casimir force in the limiting case of extremely large plates
separation,

\begin{equation}
f=\frac{\Gamma(-2)}{8}(\sum_{l=0}^{d-1}\left(%
\begin{array}{c}
  d \\
  l \\
\end{array}%
\right)E_{d-l}(1,1,\cdot\cdot\cdot,1;-2))\frac{1}{L^{4}}
\end{equation}

\noindent Here the dimensionality of the high dimensional
spacetime is $D=4+d$, $d$ the number of extra compactified
dimensions. The burden and surprisingly difficult calculation is
performed to regularize (7), then the asymptotic value of Casimir
force for parallel plates in the spacetime with $d$ extra
dimensions is obtained and depicted in Figure 1. In the spacetimes
with different dimensionality, the asymptotic value of Casimir
force for parallel plates in the limiting of extremely large
plates distance is definite and nonnegative. When the
dimensionality is four, the value of Casimir force vanishes in the
limiting case. We find that the asymptotic values are positive in
the background with more than four dimensions, which means that
there must exist repulsive Casimir force as the plates gap is
large enough in the presence of additional compactified dimensions
no matter how long the size of extra dimensions is. It is
interesting that the asymptotic values of Casimir force depend on
the dimensionality of spacetime, the higher dimensionality, the
greater asymptotic value. We should not neglect that the repulsive
Casimir force between parallel plates is excluded in the practice.
We must point out that the experiment is always performed on
electromagnetic fields that may obey more complicated boundary
conditions than the case of scalar field we consider here, and the
asymptotic value of Casimir force for different kinds of fields
satisfying different boundary conditions will certainly be
different, but all of the asymptotic values keep positive, which
means that the repulsive Casimir force must appear as the plates
are sufficiently far away from each other.

In conclusion, the model of higher dimensional spacetime described
by standard Kaluza-Klein theory can not be realistic. Having
discussed the Casimir force between parallel plates in the frame
of Kaluza-Klein approach in detail, we discover that the values of
Casimir force always remain positive as the plates move apart from
each other to farther enough, which means that there must exist
the repulsive Casimir force when the separation is sufficiently
large. The experimental evidence confirm that no repulsive Casimir
force appear in this case. Although we are limited here by only
the asymptotic case for simplicity and comparison, it is enough
for us to declare that the results obtained from the standard
Kaluza-Klein theory including extra compactified dimensions
disagree with the experimental evidence inevitably, leading a
reliable negative verdict on the theory. The Kaluza-Klein theory
needs to be developed further and related topics also need further
research.

\vspace{3cm}

\noindent\textbf{Acknowledgement}

The author thank Professor I. Brevik and Professor E. Elizalde for
helpful discussions. This work is supported by the Shanghai
Municipal Science and Technology Commission No. 04dz05905.

\newpage
\begin{figure}
\setlength{\belowcaptionskip}{10pt} \centering
  \includegraphics[width=15cm]{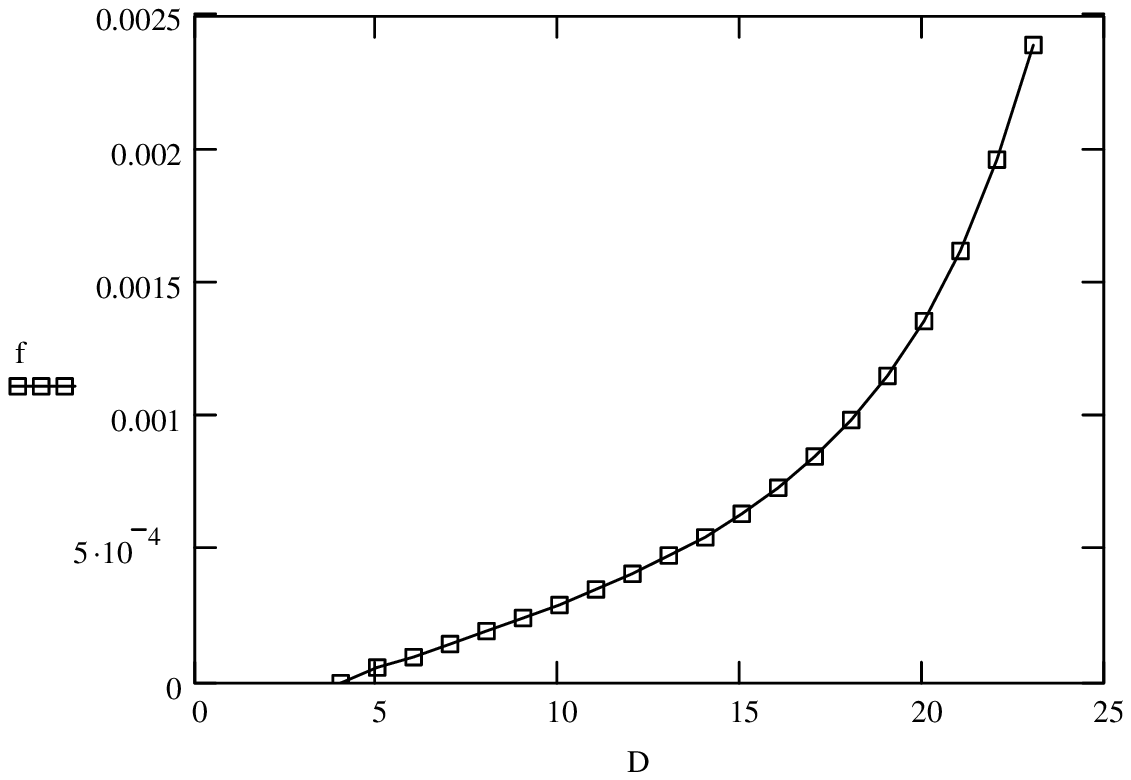}
  \caption{The Casimir force for parallel plates versus dimensionality of spacetime as the separation of plates goes to the infinity.}
\end{figure}

\end{document}